\documentclass[conference]{IEEEtran}
\IEEEoverridecommandlockouts
\usepackage{cite}
\usepackage{amsmath,amssymb,amsfonts}
\usepackage{gensymb}
\usepackage{cuted}
\usepackage{multicol}
\usepackage{multirow}
\usepackage{makecell}
\usepackage{capt-of}
\usepackage{algorithmic}
\usepackage{graphicx}
\usepackage{textcomp}
\usepackage{xcolor}
\usepackage{indentfirst}
\usepackage{ragged2e}
\usepackage{flushend}
\usepackage{booktabs}
\usepackage{etoolbox} % for \robustify
\usepackage[caption=false,font=footnotesize]{subfig}
\usepackage{stfloats}
\usepackage{setspace}
\usepackage[font=footnotesize,labelsep=period]{caption}

\begin{document}

%\title{Experimental Evaluation of Thickness and Radial Mode Power Transfer Capability in Through-Metal Acoustic Power Transfer
\title{Experimental Characterization and Prediction of Radial and Thickness Mode Power Transfer Capability in Through-Metal Acoustic Power Transfer

\thanks{
Research was sponsored by the Army Research Laboratory and was accomplished under Cooperative Agreement Number W911NF2220007. The views and conclusions contained in this document are those of the authors and should not be interpreted as representing the official policies, either expressed or implied, of the Army Research Office or the U.S. Government. The U.S. Government is authorized to reproduce and distribute reprints for Government purposes notwithstanding any copyright notation herein.}
}

\author{

\IEEEauthorblockN{
Anwesha Mukhopadhyay\textsuperscript{1},
Avinash Arya\textsuperscript{2},
Daniel Costinett\textsuperscript{2},\\
Sarah S. Bedair\textsuperscript{3},
Victor Farm-Guoo Tseng\textsuperscript{3}
}

\IEEEauthorblockA{\textsuperscript{1}
Department of Electrical Engineering, IIT Bombay, Mumbai, India\\
anwesham@iitb.ac.in
}

\IEEEauthorblockA{\textsuperscript{2}
Department of Electrical Engineering and Computer Science\\
The University of Tennessee, Knoxville, TN, USA
}

\IEEEauthorblockA{\textsuperscript{3}
U.S. Army Combat Capabilities Development Command
Army Research Laboratory, USA
}

}
\maketitle

\begin{abstract}
Through-metal acoustic power transfer (TM-APT) is an emerging wireless solution for battery charging in hazardous and sensitive environments where electronics are enclosed within sealed metal structures without electrical feed-through. Power transfer is achieved utilizing reverse and direct piezoelectric effect via acoustic wave propagation through the metal barrier. The achievable power, often targeting hundreds of watts, is strongly influenced by different factors, such as, material properties, geometry, dimensions of the metal and piezo, and their bonding, making maximization of power transfer capability critical.
This work experimentally investigates the terminal electrical characteristics of multiple prototypes, representing realistic scenarios. It compares power transfer capability in thickness and radial resonance modes for piezo–metal assemblies, enabling appropriate mode selection and power electronics interface design. A figure of merit (FOM) is proposed for rapid screening and ranking of configurations as far as the power transfer capability is concerned.
\end{abstract}

\noindent\textbf{\textit{Keywords-Acoustic wireless power transfer, Piezo resonator, Radial mode, Thickness mode, Power transfer figure of merit}}

\section{Introduction}
Acoustic power transfer has gained significant attention over the years for applications ranging from powering sensors, defect detection systems\cite{defectdetection}, exploration of marine resources \cite{Underwater} to wearable electronics and biomedical implants \cite{implants}, covering a wide range of power levels. The underlying principle of power transfer relies on the piezoelectric and reverse piezoelectric effects. Electrical energy applied to a piezoelectric resonator is converted into vibrational energy, which propagates through a medium as acoustic waves, typically in the ultrasonic frequency range. At the receiver, the vibrational energy is subsequently converted back into electrical energy and utilized for applications after appropriate power conditioning.

Piezoelectric resonators capable of handling several kilowatts of power have been reported in the literature \cite{Spuriousfree}, where customized electrode configurations suppress spurious modes and maximize power transfer capability. A piezoelectric-transformer-based DC-DC converter \cite{PiezoTransformer} has demonstrated high power-transfer efficiency with a single piezo and bulls-eye electrode configuration on both faces, acting as the primary and secondary of the transformer, but the reported power level is below 10 W.
\begin{figure} [b!] \vspace{-10pt}
    \centering
     %\subfloat[]
    {\includegraphics[width=1\linewidth]{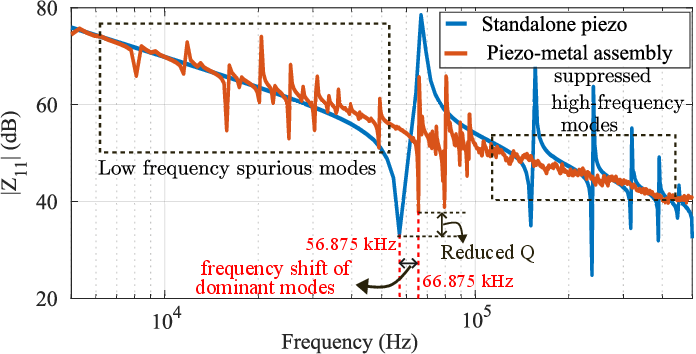}} 
   %  \subfloat[]
   % {\includegraphics[width=0.5\linewidth]{Figures/OptimizationFlowChart.eps}}
    \caption{Deviation in impedance characteristics from standalone piezo to piezo–metal assembly showing reduced Q-factor, resonance frequency drift, and spurious mode generation.}
    \label{fig:stanalonevsassembled}
\end{figure}
In through-metal acoustic power transfer, however, the metal barrier serves as the coupling medium, requiring separate transmitter (Tx) and receiver (Rx) piezoelectric elements bonded to two opposite sides of the metal. As a result, the characteristics of the resulting piezo–metal assembly can differ significantly from those of a standalone piezoelectric resonator. The presence of adhesive and constrained displacement at piezo-adhesive-metal boundaries can lead to a reduction in quality factor and the appearance of spurious modes, which distribute the power transfer capability over a range of frequencies rather than concentrating it at a narrow band of frequency. Methods to suppress spurious modes through electrode shape modification \cite{Spuriousfree,spuriousfree2} and improved mechanical fixtures \cite{mechfixture}, explored for single piezo resonators, are often impracticable to adopt in TM-APT due to restricted electrode accessibility, compromised piezo-metal bonding strength, and reliability of the electrical connections. Additionally, manual assembly processes, including surface preparation, adhesive application, curing, and wire attachment introduce additional variations among prototypes.
The deviation in electrical characteristics between a standalone piezoelectric element and a corresponding  piezo–metal assembly is illustrated in Fig. \ref{fig:stanalonevsassembled}, highlighting the following:
\begin{enumerate} 
    \renewcommand{\labelenumi}{(\roman{enumi})}
    \item shift in the dominant resonance frequency,
    \item reduced separation between resonance and antiresonance frequencies,
    \item increased damping of the dominant resonance mode, and
    \item generation of lower-order spurious modes.
\end{enumerate}
\begin{figure*}[b!] \vspace{-10pt}
   \centering
   %   \subfloat[]
   %   {\includegraphics[width=0.45\columnwidth]{Figures/AcousticCkt2.eps}} 
      % \subfloat[]
     {\includegraphics[width=0.9\textwidth]{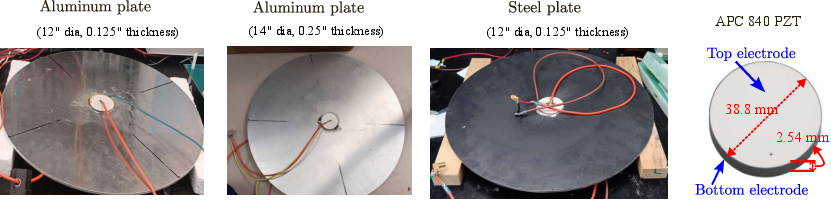}} \\
     % \subfloat[]
     % {\includegraphics[width=0.85\columnwidth]{Figures/wavelengthandplatethickness2.eps}}
     \caption{Different setups with metal plate and disc-type piezo.}
     \label{fig:setupandckt}
\end{figure*}

These effects collectively reduce power transfer capability by lowering modal Q-factor, altering the optimal operating frequency, and distributing power transfer possibility among multiple resonances. While lower-order spurious modes do not contribute significantly to useful power transfer under the intended operating frequency, suppressed higher-order modes reduce the possibility of leveraging alternative modes for power transfer at the harmonic frequency.
Therefore, selection of the appropriate excitation frequency, voltage, and loading arrangement is essential for maximizing power transfer, while not violating the temperature and depolarization limits of the piezo resonator.

Conventional piezoelectric resonator models, such as the KLM and Mason formulations \cite{2d_KLM}, generally do not capture assembly-induced non-idealities and variations. While these models can often predict the overall behavior of thickness-mode operation with reasonable accuracy, larger discrepancies may arise for radial-mode operation due to stronger sensitivity to adhesive properties, plate geometry, boundary conditions, and assembly variations. Consequently, the predicted resonance characteristics and power transfer capability can deviate significantly from experimental observations, particularly in practical piezo–metal assemblies.
 % where maximizing power transfer while adhering to thermal and depolarization limits requires proper selection of operating frequency, drive voltage, and loading conditions.
 
To address these challenges, an experimental approach is adopted in this work to obtain the frequency-dependent electrical characteristics of piezo–metal assemblies. These characteristics are used to construct a two-port network at each frequency over the range of interest. Maximum achievable power is determined by identifying the global optimum under practical constraints of depolarization voltage and Curie temperature limits. Furthermore, a figure of merit (FOM) is proposed to enable rapid screening and ranking of prototypes based on power transfer capability without requiring detailed network solutions. Practical challenges associated with characterization, data anomalies, assembly variability, and nonlinear operation that can affect power-transfer prediction are also discussed.

The remainder of the paper is organized as follows. Section~\ref{sec:SecII} describes the experimental setup, prototype building, and criteria used for selecting piezoelectric elements and metal plates, and compares the impedance characteristics of thickness and radial modes obtained using a vector network analyzer (VNA). Section~\ref{sec:SecIII} presents the two-port optimization framework for determining optimal operating conditions and introduces the proposed FOM. Section~\ref{sec:SecIV} discusses practical measurement considerations, limitations of the characterization process, and their impact on power-transfer prediction, followed by the conclusions.

\section{Through Metal Acoustic Power Transfer Set-up and Characterization} \label{sec:SecII}

% Fig. 1(a) depicts the acoustic power transfer setup, consisting of a metal body to which transmitter(Tx) and receiver (Rx) piezo discs are attached. 
% piezoelectric discs are attached. One of the piezo discs, referred as the transmitter (Tx) piezo, is excited from a voltage $v_{tx}$ to produce vibration by inverse piezoelectric effect. The vibrational leads to vibration or flexural bending of the plate, resulting in strain on the other side piezo disc, referred as receiver (Rx) piezo. Due to direct piezoelectric effect, the receiver piezo generates a voltage $v_{rx}$, which is processed through a power converter to charge the battery. 
Fig. \ref{fig:setupandckt} shows multiple laboratory-scale prototypes mimicking the scenario of piezo attachment to a sealed metal container.  Circular aluminum and steel plates are used, with piezo discs bonded at the center of each face with silver epoxy adhesive. 
%The conductive adhesive keeps the bottom electrodes at the same potential as the metal plate, enabling electrical access without direct wiring, facilitating uniform adhesive contact across the bonding interface.
% Non conducting adhesive with wrap around or bull's eye electrode configurations are other combinatorial possibilities for this setup. 
% The plate and piezo dimensions and materials are listed in Table~I.
% \usepackage{multirow}
 \begin{table*}[!b]
\centering
\caption{Piezo Properties and Calculation of Plate Thickness}
\begin{tabular}{ccclcl}
\hline
\multicolumn{1}{|c|}{\multirow{2}{*}{Piezo}} & \multicolumn{1}{c|}{\multirow{2}{*}{\begin{tabular}[c]{@{}c@{}}Diameter\\ $d_p$\end{tabular}}} & \multicolumn{1}{c|}{\multirow{2}{*}{\begin{tabular}[c]{@{}c@{}}Thickness\\ $h_p$\end{tabular}}} & \multicolumn{1}{c|}{\multirow{2}{*}{\begin{tabular}[c]{@{}c@{}}Density\\ $\rho_p$\end{tabular}}} & \multicolumn{2}{c|}{Frequency  Constant (m/s)}                               \\ \cline{5-6} 
\multicolumn{1}{|c|}{}                          & \multicolumn{1}{c|}{}                                                                          & \multicolumn{1}{c|}{}                                                                              & \multicolumn{1}{c|}{}                                                                          & \multicolumn{1}{c|}{Thickness ($N_T$)} & \multicolumn{1}{c|}{Planar ($N_p$)} \\ \hline
\multicolumn{1}{|c|}{PZT-APC 840}               & \multicolumn{1}{c|}{38.8 mm}                                                                   & \multicolumn{1}{c|}{2.54 mm}                                                                       & \multicolumn{1}{c|}{7400 kg/m$^3$}                                                             & \multicolumn{1}{c|}{2005}              & \multicolumn{1}{c|}{2130}           \\ \hline \hline

\multicolumn{1}{|c|}{Metal} & \multicolumn{1}{c|}{Young's Modulus Y} & \multicolumn{1}{c|}{Density $\rho_m$}  & \multicolumn{1}{l|}{Velocity $v=\sqrt{\frac{Y}{\rho_m}}$}   & \multicolumn{2}{c|}{Metal Thickness $h_{m}=v/2f_r$}              \\ \hline               \multicolumn{1}{|c|}{Aluminum} &\multicolumn{1}{c|}{$7.03\times 10^{10}$ Pa}  & \multicolumn{1}{c|}{2670 kg/m$^3$} &\multicolumn{1}{c|}{5131.24 m/s}                                                               &\multicolumn{2}{c|}{3.25 mm} \\
\multicolumn{1}{|c|} {Steel} &\multicolumn{1}{c|}{$21\times 10^{10}$ Pa} & \multicolumn{1}{c|}{7870 kg/m$^3$} & \multicolumn{1}{c|}{5165.62 m/s}                                                               & \multicolumn{2}{c|}{3.27 mm}
\\ \hline     
\end{tabular}
\label{tab:properties&dimension}
 \end{table*}
 \subsection{Selection of Piezoelectric Material and Dimension}
High power transfer requires application of higher voltage and higher electromechanical power conversion efficiency. In order to withstand higher applied voltage, the dimension of the piezo resonator should be appropriately selected so that the maximum field stress is not exceeded to drive the piezo into depolarization. A hard piezoelectric material APC 840 PZT is selected due to its low dielectric losses and high mechanical quality factor, which ensures higher power conversion efficiency and thus better thermal management. Considering the maximum allowable electric field along the thickness direction ($E_3$) as 9–11 VAC/mil  \cite{APC840_properties}, the thickness of both transmitter and receiver piezoelectric elements is selected as 2.54 mm, enabling operation up to approximately 900 VAC. Additionally, the higher coercive field of the selected piezo material improves resilience against depolarization under high-voltage operation. Diameter of 38.8 mm, the largest available off-the-shelf size, is selected to enhance heat dissipation and is consistent with the dimensions reported in high-power acoustic applications \cite{highpower}.
The dimensions and relevant material properties of the piezo disc are provided in Table~\ref{tab:properties&dimension}. 
% While low dielectric loss and high quality factor help limit losses and temperature rise under higher power operation, a high coercive field improves resilience against depolarization during higher excitation voltage.
% In addition to material selection, the piezo dimensions are chosen to withstand high excitation voltage while providing sufficient surface area for heat dissipation, maintaining the operating temperature well below half of the Curie temperature. The dimensions and relevant material properties of the piezo disc are provided in Table I. With a maximum allowable electric field between 9–11 VAC/mil for APC 840 \cite{APC840_properties}, a thickness of 2.54 mm enables operation up to approximately 900 VAC. A diameter of 38.8 mm, the largest available off-the-shelf size, is selected to enhance heat dissipation and is comparable to those used in high-power acoustic applications \cite{highpower}.

\subsection{Metal Plate Geometry and Dimensions}
The geometry and size of the metal plate are selected keeping an eye on two purposes, viz., the plate geometry must be a reasonable approximation of a sealed metal enclosure and the mechanical resonance frequency of the plate is close to the piezo self-resonance for efficient acoustic coupling. A circular plate is selected as a finite section of a large metallic enclosure. Prior work shows that a square plate \cite{Kody} or even a square-shaped \cite{Spuriousfree} piezo generates a large number of spurious modes, making selection of excitation frequency difficult, hence not considered in this work. Thickness of the metal plate is selected to create mechanical resonance such that the self-resonant frequency of the piezo disc and the metal plate coincide. The thickness frequency constant $N_T$ for APC 840 results in a resonance frequency of $f_t = N_T/h_p = 789.37$ kHz for the selected 2.54 mm thick piezo disc. 
 For mechanical resonance, calculations of the plate thicknesses are shown in Table~\ref{tab:properties&dimension}, considering nodes and antinodes of the propagating wave as depicted in Fig.~3. Since the assumption of mechanical resonance does not account for the presence of the adhesive layer between the piezo and the metal plate, slight deviations in the resonance frequency and higher damping may result depending on the thickness and stiffness of the adhesive. However, as long as the adhesive thickness is much smaller than the acoustic wavelength within the adhesive, its influence on the resonance frequency is expected to be relatively small, and the mechanical resonance approximation remains reasonable. Experiments use three prototypes of the following dimensions:
 \begin{itemize}
 \renewcommand{\labelitemi}{--}
     \item I. Al plate of 12~in dia and 0.125~in (3.175mm) thickness
     \item II. Al plate of 14~in dia and 0.25~in (6.35mm) thickness
     \item III. Steel plate of 12~in dia and 0.125~in (3.175mm) thickness
 \end{itemize}
 The radial-mode self-resonating frequency of the piezo is calculated as $f_r=N_p/d_p=54.89$ kHz. Bonding the piezo to the metal plate introduces additional flexural modes that typically appear at frequencies below the radial resonance frequency of standalone piezo. The plate diameter is chosen around ten times the piezo diameter and is not decided from mechanical resonance considerations. An appropriately chosen plate dimension can suppress the lower flexural mode frequencies and maximize power transfer capability at the piezo self-resonance. 
 \begin{figure}[t!]
    \centering
     {\includegraphics[width=0.9\columnwidth]{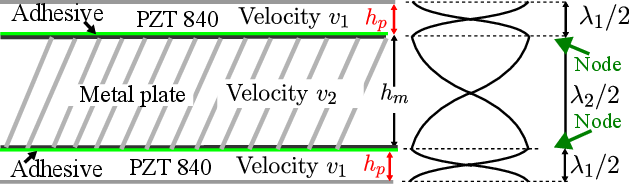}}
    \caption{Cross-section of a plate and ideal propagation of acoustic wave in thickness mode creating nodes and antinodes.}
    \label{fig:placeholder} \vspace{-10pt}
\end{figure}
\subsection{Surface Preparation and Attachment of Piezo}
The metal surface is cleaned and roughened with sandpaper to ensure a mechanically reliable and uniform attachment. Two-part silver epoxy 8331D, mixed in 1:1, is uniformly applied at the center of the plate, covering an area equal to that of the piezo. The assembly is put under vacuum for 30 mins and subsequently cured at room temperature for more than six hours.
The conductive adhesive ensures electrical continuity between the bottom electrode and the metal plate, while also providing good thermal conduction, facilitating utilization of the metal plate for transferring heat from the piezo. Electrical access to the bonded face through the metal plate without direct wiring ensures uniform adhesive contact across the bonding interface. Maintaining a uniform and optimal adhesive thickness \cite{Meesala2021UAET} across all prototypes is critical for fair comparisons. 
It is also to be noted that by using conductive adhesive, the galvanic isolation between the Tx and Rx piezo is compromised. Since the conductive adhesive electrically connects both piezos through the metal plate, non-isolated probing can introduce circulating currents at high-frequency excitation, resulting in apparent efficiency degradation, especially in the thickness mode. Therefore, care should be taken to monitor potential common-mode current flow. If necessary,  common-mode choke should be employed to suppress the common-mode current and minimize measurement errors.
\begin{figure*}[t!] 
  \centering
   %   \subfloat[]
   %   {\includegraphics[width=0.45\columnwidth]{Figures/AcousticCkt2.eps}} 
      \subfloat[]
     {\includegraphics[width=0.95\columnwidth]{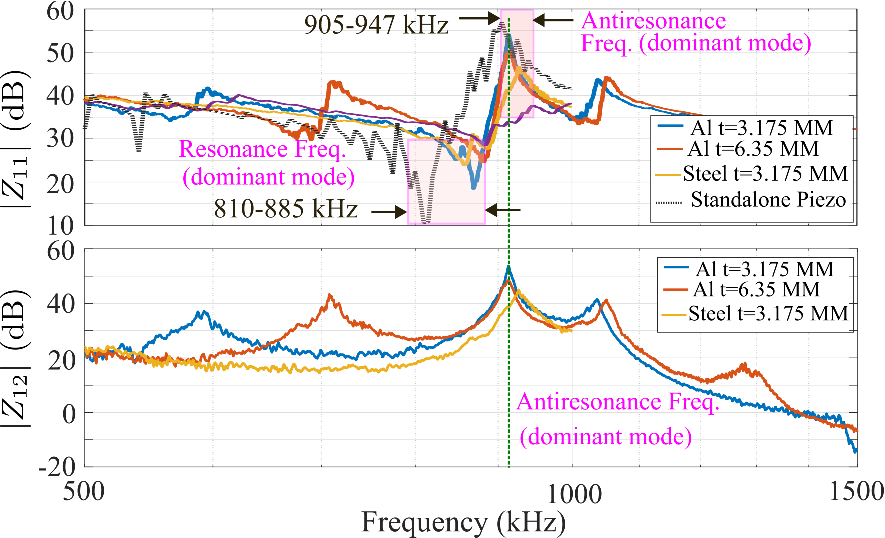}} \qquad
     \subfloat[]
     {\includegraphics[width=0.95\columnwidth]{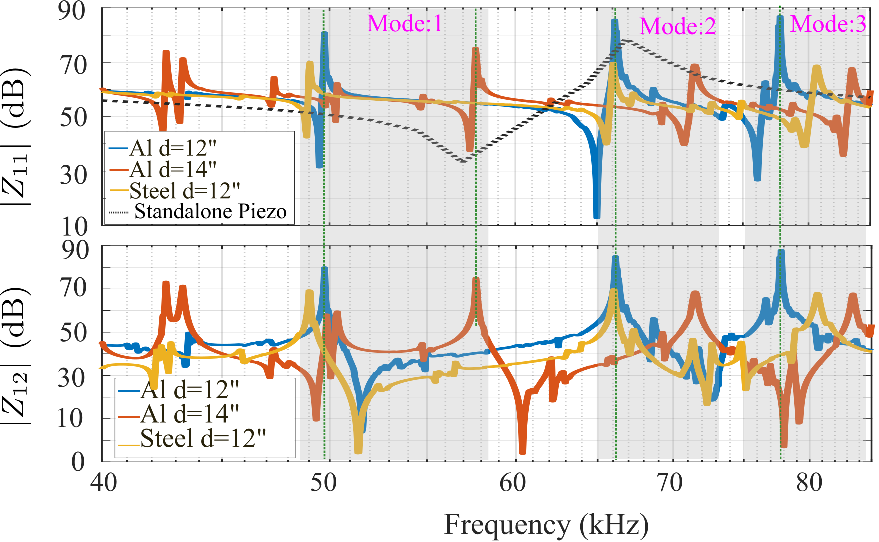}}      
     \caption{$Z_{11}$ and $Z_{12}$-characteristics captured for frequencies corresponding to (a) thickness and (b) radial modes for different prototypes.}    
  \label{fig:PiezoModes} \vspace{-10pt}
\end{figure*}
\subsection{Characterization of Thickness and Radial Modes}
With the Vector Network Analyzer (VNA) E5061B, S-parameters of different prototypes are measured. For higher resolution, the number of data points is set to maximum (1601), and IF bandwidth is set to 10 Hz for better accuracy while compromising on the characterization speed. A poor resolution with fewer data points and higher IF bandwidth, both can cause measurement inaccuracy, especially in high Q resonances.  The captured S-parameters are subsequently converted to Z-parameters using MATLAB functions. The impedance $|Z_{11}|$ and $|Z_{12}|$ variations over frequency for the dominant thickness mode and radial modes between 40 kHz to 85 kHz are depicted in Figs.~\ref{fig:PiezoModes}(a) and (b), respectively. 
The overlaid standalone piezo characteristics indicate a significant reduction in Q-factor and about 10$\%$-shift in the resonance frequency in the thickness mode for all the prototypes. The radial mode, however, depicts multiple significant spurious modes, based on the planar dimensions of the plates, while the change in Q-factor is negligible. The separation between resonance and anti-resonance frequencies, which is affected by the adhesive layer, appeared significantly reduced in the radial mode compared to the standalone piezo, making optimal frequency control challenging.

\begin{figure}[b!] \vspace{-10pt}
  \centering
          {\includegraphics[width=0.55\columnwidth]{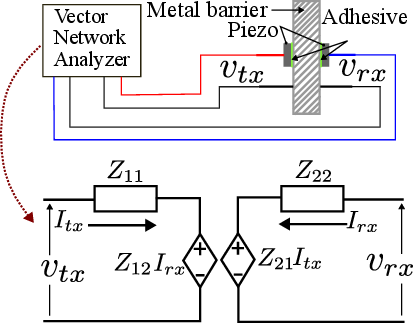}  }
     \caption{Equivalent two-port network using Z-parameters from VNA.}    
  \label{fig:twoport}
\end{figure}

\begin{figure}[b!] \vspace{-10pt}
  % \centering
   %   \subfloat[]
   %   {\includegraphics[width=0.45\columnwidth]{Figures/AcousticCkt2.eps}} 
      \subfloat[]
     {\includegraphics[width=0.24\textwidth]{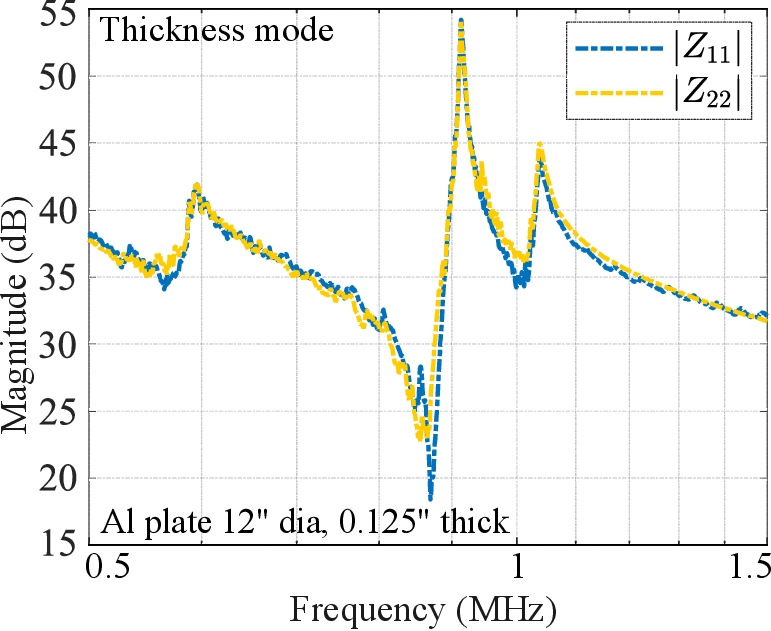}} 
      \subfloat[]
     {\includegraphics[width=0.235\textwidth]{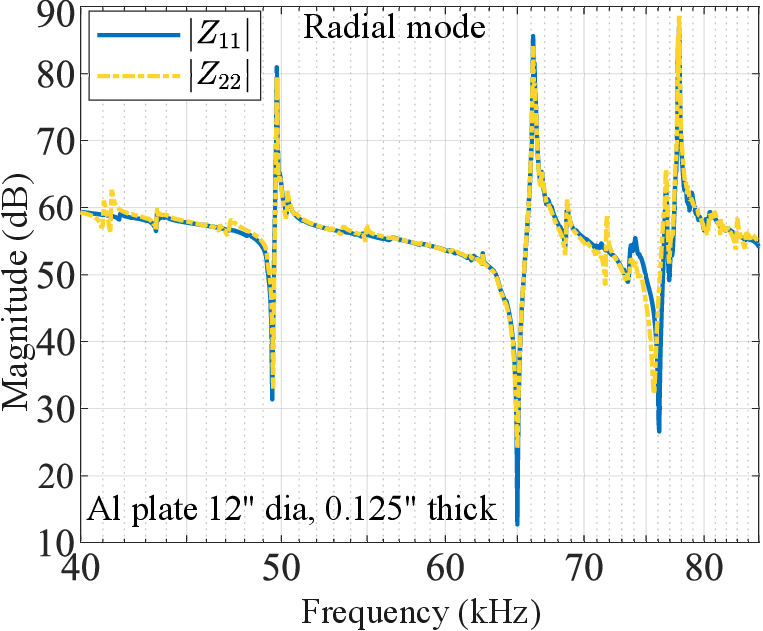}}\\
     \subfloat[]
     {\includegraphics[width=0.24\textwidth]{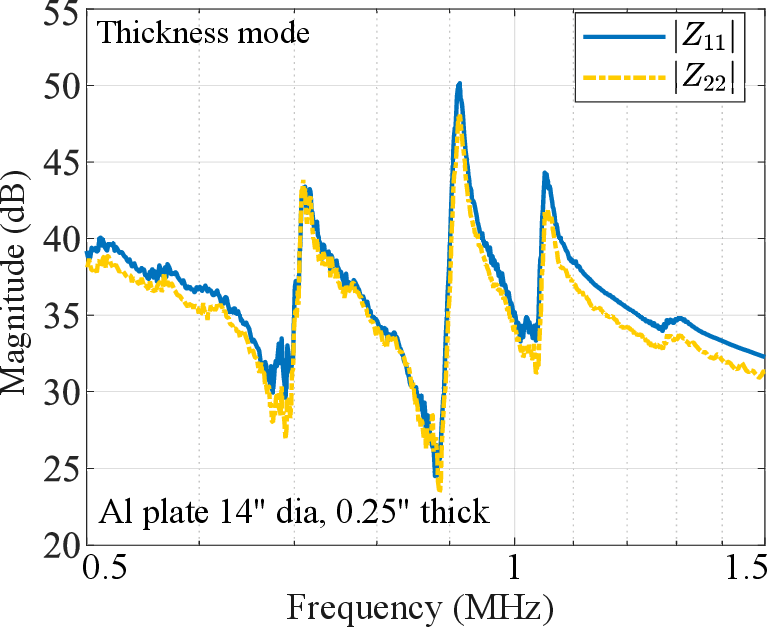}} 
     % \subfloat[]
     % {\includegraphics[width=0.3\textwidth]{Figures/Z11Z22_Steel12INCHThickness.eps}} \\
         \subfloat[]
     {\includegraphics[width=0.235\textwidth]{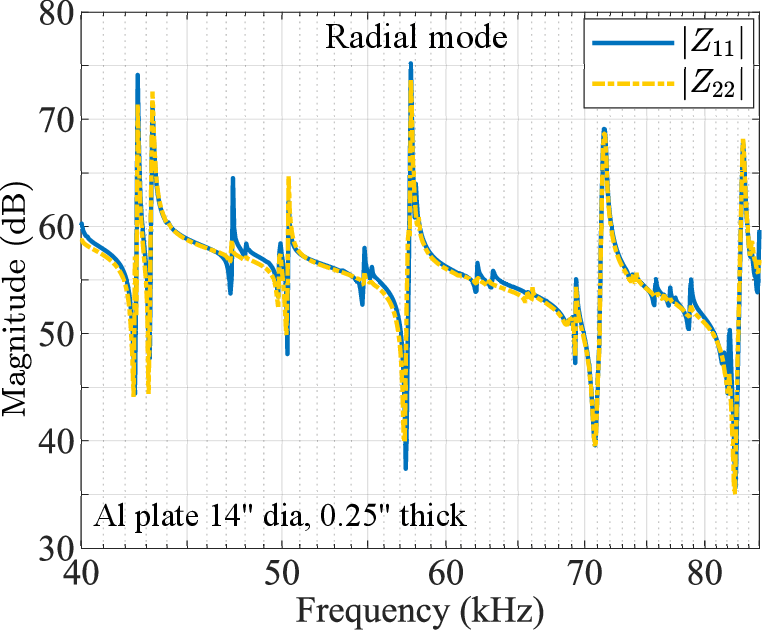}} 
    % \subfloat[]
     % {\includegraphics[width=0.3\textwidth]{Figures/Z11Z22_Steel12INCHRadial.eps}} 
          \caption{Overlaid $Z_{11}$ and $Z_{22}$ thickness and radial mode characteristics for (a) and (b) 12 inch dia Al plate, (c) and (d) 14 inch dia Al plate, respectively.}    
  \label{fig:nonsymmetric}
\end{figure}
The distribution of different radial modes across three setups shows that plates with the same diameter exhibit modes at the same frequencies, as acoustic wave velocity is nearly equal in Al and steel; however, the larger-diameter plates tend to shift the frequencies further to the right side of the scale. It can be seen that despite having the same dimensions of plate and identical piezo and adhesive material, the 12~in dia Al plate produces high-Q resonance at equal or near-equal resonance frequencies compared to that of the steel plate. The Z-parameters ($Z_{11}$, $Z_{12}$, $Z_{21}$, and $Z_{22}$) obtained from the S-parameters captured by VNA are used to construct an equivalent two-port network, as shown in Fig.~\ref{fig:twoport} at each frequency, captured in the range of interest. Although the network is theoretically symmetric ($Z_{11}=Z_{22}$) and reciprocal ($Z_{12}=Z_{21}$), practical factors such as bonding variations, wire attachment, and measurement uncertainties introduce asymmetry between $Z_{11}$ and $Z_{22}$. Fig.~\ref{fig:nonsymmetric} depicts $Z_{11}$ and $Z_{22}$, highlighting the mismatch in Q-factors at the resonance frequencies of the Al-plate prototypes in both the thickness and radial modes. 
Despite identical VNA measurement settings, the observed mismatch in the Q-factor indicates variations in the adhesive layer and wire attachment between the Tx and Rx piezoelectric transducers. Such assembly-induced variations can significantly affect measurement-based power-transfer prediction, highlighting the need for a standardized assembly and characterization procedure. The transfer impedances $Z_{12}$ and $Z_{21}$ remain indistinguishable for all the cases considered, hence are not shown here.
%More discussion on all other Z- parameters and spurious radial-modes will be included in the full paper. 
%measured using Keysight E5061B vector network analyzer,
% different changes across the setups. Th The shows slight deviation in  dominant resonant frequencies both in thickness and radial modes from that of the calculated values. The deviation can be attributed to tolerance in parameter and dimension and attachment of wire at the center of both faces, which apparently affects the radial mode frequency negligibly (0.23$\%$) compared to the thickness (2.6$\%$). 
% \begin{figure}
%     \includegraphics[width=1\linewidth]{Figures/thicknessModeZplots.eps}
%  \end{figure}
% The post metal attachment impedance characteristics for thickness and radial modes are depicted in Fig. 3(a) and (b).
% Fig. 2(a)  show the $Z-$ parameters $Z_{11}$ and $Z_{12}$ for  different plate assemblies focusing on the thickness mode resonance. Percentage deviation in the resonance frequencies are.... . 

\section{Power Transfer Calculation and Proposed Figure of Merit}  \label{sec:SecIII}
\subsection{Power Transfer and Optimal Operation}
\begin{figure} [t!] \vspace{-10pt}
     \centering
     \subfloat[] 
     %{\includegraphics[width=1\linewidth]{Figures/shrinkFlowChart.eps}}\\ 
   {\includegraphics[width=0.9\columnwidth]{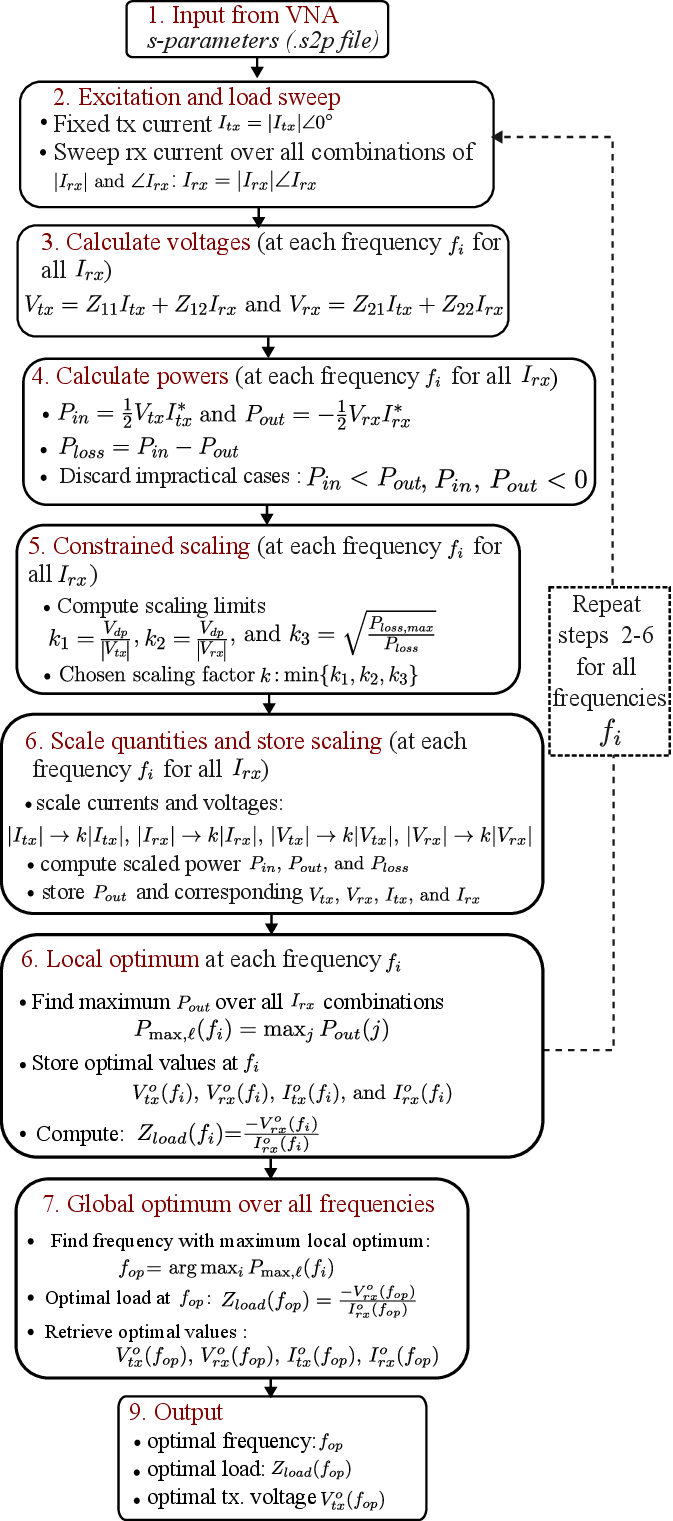}}  
   \caption{Flow-chart depicting steps to find optimal operating point from characterization data obtained with VNA.}
 \label{fig:flowchart} \vspace{-10pt}
\end{figure}

The maximum power transfer is evaluated from the two-port network at each frequency of the VNA sweep. For a fixed input current phasor $I_{tx}$ and a practical range of output current sweep, the transferred power is computed under depolarization voltage ($V_{dp}$) and thermal constraints. The flow-chart in Fig.~\ref{fig:flowchart} outlines the procedure to determine the optimal operating frequency $f_{op}$, load $Z_{load}(f_{op})$, and applied voltage to the Tx piezo $V_{tx}(f_{op})$. 

Table~\ref{tab:OptPoints} presents the operating voltage, frequency, and optimal load to achieve maximum power transfer for all three prototypes for both thickness and the most dominant radial mode. The optimal frequency of operation usually lies between the resonance ($f_r$) and anti-resonant ($f_{ar}$) frequencies; however, the depolarization and thermal constraints can shift the optimum slightly to below $f_r$ or above $f_{ar}$. As shown in Table~\ref{tab:OptPoints}, the dominant radial mode for all the prototypes give $f_{op}<f_r$, whereas, in thickness mode aluminum plates satisfy $f_r<f_{op}<f_{ar}$ while the steel plate yields $f_{op}>f_{ar}$. 

The flowchart in Fig.~\ref{fig:flowchart} determines the optimal complex load impedance, $Z_{load}(f_{op})$, whose imaginary component is mentioned as the Rx-side compensation in Table~\ref{tab:OptPoints}. The required compensation, inductive or capacitive, depends on the relative phase relationship between $Z_{11}$ and $Z_{12}$ at the optimal operating frequency. For all three prototypes and operating modes, inductive compensation is required when ($f_{op}<f_r$) or ($f_{op}>f_{ar}$), whereas capacitive compensation is required when ($f_r<f_{op}<f_{ar}$). An exception is observed in the thickness mode of the prototype~II, where the presence of two closely spaced thickness modes alters this trend.

The comparison shows that the predicted power transfer capability for the radial modes is higher for all the prototypes; however, it requires higher equivalent load resistance and inductive compensation to achieve that. The operating point calculation considers that maximum peak voltages applied across the Tx or Rx are less than 520 V, which is about 50$\%$ of the depolarization voltage $v_{dp}$, and power loss is less than 45 W, calculated using the thermal model discussed in \cite{Kody} to limit the maximum piezo temperature below 100 $^\circ$C.
The estimated optimal power in thickness and the most significant radial mode is shown in Fig.~\ref{fig:FOM} (a),~(b)(i)--(iii) for three prototypes, overlaid on their respective $|Z_{11}|$.
\begin{table*}[tb!]
\centering
\caption{Maximum Power and Optimal Operating Points}
\begin{tabular}{c| c| c c c c c}
\hline
\multirow{2}{*}{Prototype}            & \multirow{2}{*}{Mode}      & \multirow{2}{*}{\begin{tabular}[c]{@{}c@{}}Maximum \\ Power (W)\end{tabular}} & \multirow{2}{*}{\begin{tabular}[c]{@{}c@{}}Optimal \\ Frequency  (kHz)\end{tabular}} &$|V_{tx}|$ (V) & \multirow{2}{*}{\begin{tabular}[c]{@{}c@{}}Optimal \\ Load ($\Omega$)\end{tabular}} & \multirow{2}{*}{\begin{tabular}[c]{@{}c@{}}Rx-end \\ Compensation\end{tabular}} \\
                                      &                            &                                                                               &                                                                                      &                                                                                     &                                                                                 \\ \hline
I   & \multirow{3}{*}{Thickness} & 67.8                                                                          & 873.5 $\in (f_r, f_{ar})$                    & 98.2                                                         & 58.2                                                                                & 3.6 nF                                                                          \\ %\cline{1-1} \cline{3-6} 
II   &                            & 77.3                                                                          & 902.4 $\in (f_r, f_{ar})$                  & 157.9                                                            & 67.9                                                                                & 5.4 $\mu$H                                                                        \\ %\cline{1-1} \cline{3-6} 
III &                            & 31.4                                                                          & 926 $>f_{ar}$                      & 115                                                          & 89.3                                                                                & 5.3 $\mu$H                                                                       \\ \hline
I   & \multirow{3}{*}{Radial}    & 667.5                                                                         & 63.65 $<f_{r}$                      & 520.9                                                           & 130.7                                                                               & 100.7 $\mu$H                                                                    \\ %\cline{1-1} \cline{3-6} 
II   &                            & 226.7                                                                         & 57.16 $<f_r$                   & 520.7                                                            & 464.3                                                                               & 238 $\mu$H                                                                         \\ %\cline{1-1} \cline{3-6} 
III &                            & 215.5                                                                         & 65.15 $<f_r$                     & 520                                                          & 281.1                                                                               & 498.2 $\mu$H                                                                    \\ \hline
\end{tabular}
\label{tab:OptPoints}
\end{table*}

\subsection{Power Transfer Figure of Merit}
%  In a standalone piezo, performance figure of merit (FOM) is calculated from the product $k_t^2Q_r$, where $k_t$ is the electromechanical coupling coefficient \cite{APC840_properties}, which is a function of resonance and anti-resonance frequencies and $Q_m$, the mechanical quality factor indicating loss. Unlike standalone piezoelectric elements, the power transfer capability of the piezo--metal assembly depends on all four impedance parameters. To capture this combined effect, a figure of merit (FOM) is defined. The FOM is defined  as
% \begin{equation}
% FOM = \sqrt{Q_{ar}{Q_{r}}}\frac{|Z_{12}|}{\sqrt{|Z_{11} Z_{22}|}}
% \end{equation}
% where $Q_r$ and $Q_{ar}$ are the quality factors at resonance and anti-resonance frequencies calculated from the impedance characteristics considering -3 dB band. Fig.~8(a) and(b)(iv) compare the FOM for the three prototypes in thickness and radial mode~I, respectively. The frequencies of peak FOM closely match the optimal frequency obtained from power calculations, enabling rapid identification of superior configurations. Both analyses indicate that the 3.175~mm aluminum plate has the highest power transfer capability. 
% The maximum power transfer theorem from the solution of Thevenin's equivalent network gives $P_{max}=\dfrac{V^2_{Th}}{R_{Th}}$. For a given voltage input $V_{Th}$ is proportional to $\frac{Z_{12}}{Z_{11}}$ and $R_{Th}$ is proportional to $\dfrac{1}{Q}$. The FOM defined in (1) is consistent with this. To account for non-symmetry, which causes $Z_{11}$ and $Z_{22}$ to be unequal geometric mean of these two is used.
\begin{figure*} [b!]
     \centering 
   \subfloat[]
    {\includegraphics[width=0.47\linewidth]{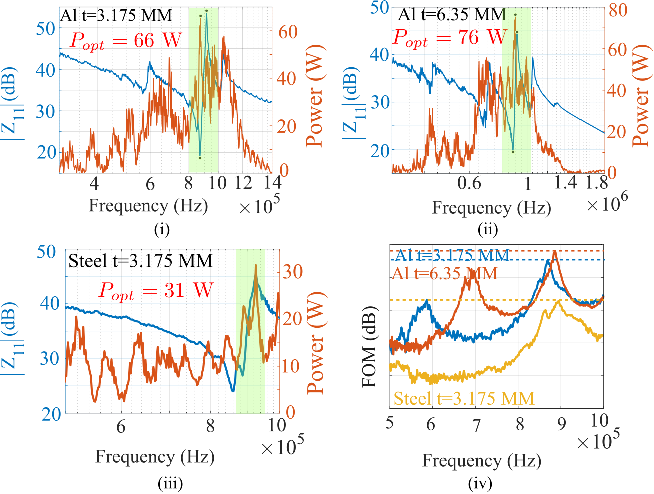}} 
   \subfloat[]
    {\includegraphics[width=0.5\linewidth]{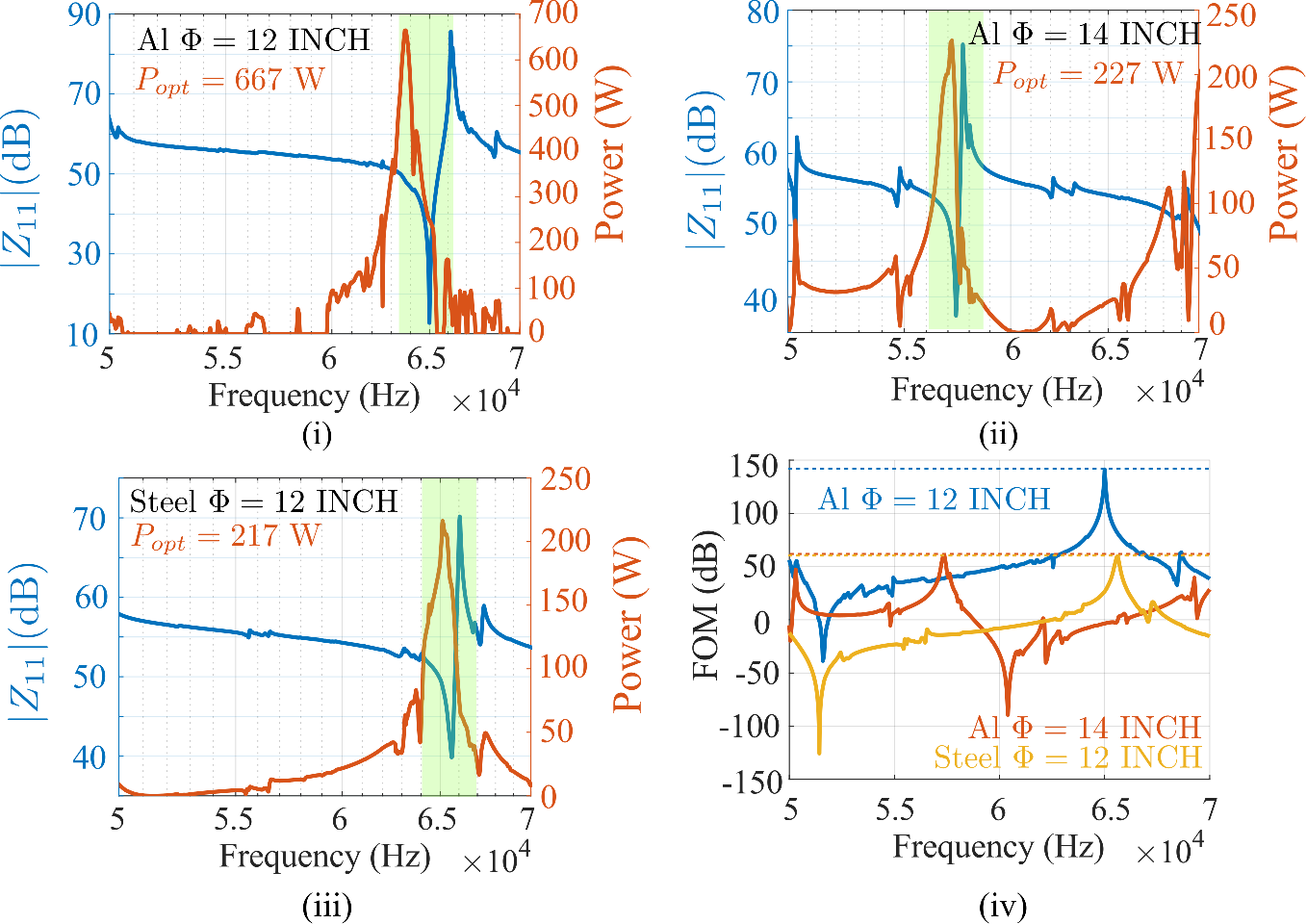}}
   %  \subfloat[]
   % {\includegraphics[width=0.20\linewidth]{Figures/thicknessmodeFOM.eps}}
       \caption{Maximum power for prototypes (i)–(iii) and FOM comparison (iv) pertaining to (a) thickness mode and (b) radial mode.}
    \label{fig:FOM}
\end{figure*}
In standalone piezoelectric resonators, the figure of merit (FOM) is commonly expressed as $k_t^2Q_m$, where $k_t$ is the electromechanical coupling coefficient and $Q_m$ is the mechanical quality factor representing resonator losses \cite{APC840_properties}. Since $k_t$ is determined by the separation between resonance and anti-resonance frequencies, the FOM combines both coupling strength and loss characteristics of the piezoelectric material. 
For the piezo--metal assembly, however, the transferable power is influenced not only by the resonance characteristics but also by the interaction between the Tx and Rx ports, represented by all four impedance parameters. To account for these effects, a figure of merit is proposed as
\begin{equation}
FOM=\sqrt{Q_{r1}Q_{r2}}\big(\frac{|Z_{12}|}{\sqrt{|Z_{11}Z_{22}|}}\big)^2
\end{equation}
where $Q_{r1}$ and $Q_{r2}$ are the quality factors evaluated at the resonance frequencies of $Z_{11}$ and $Z_{22}$, respectively, using the -3 dB bandwidth. The term $\sqrt{Q_{r1}Q_{r2}}$ is an indicator of resonant energy storage and loss, while $|Z_{12}|$ represents the strength of acoustic coupling between the Tx and Rx. The driving-point impedances $Z_{11}$ and $Z_{22}$ are included in the denominator to account for the loading and excitation characteristics of the assembly. Since practical prototypes are not perfectly symmetric, the geometric mean $\sqrt{|Z_{11}Z_{22}|}$ is used instead of either impedance individually. Figs.~\ref{fig:FOM}(a)(iv) and 8(b)(iv) compare the proposed FOM for the three prototypes in thickness and radial modes, respectively. The ranking of the prototypes based on the proposed FOM agrees with that obtained from the detailed power-transfer calculations, and the frequencies corresponding to the peak FOM closely coincide with the optimal operating frequencies.
% , enabling rapid identification of superior configurations. 
Both analyses identify the 3.175 mm aluminum plate as the prototype with the highest power-transfer potential in the radial mode. The proposed FOM is inspired by the maximum power transfer relation of the Thevenin equivalent circuit, $P_{max} \propto
V_{Th}^2/R_{Th}$. For a fixed transmitter excitation, the Thevenin voltage is proportional to the voltage transfer ratio ($Z_{12}/Z_{11}$), while higher resonator quality factor is indicative of lower effective resonator losses and thus small $R_{Th}$. Consequently, the proposed FOM combines the dominant factors affecting transferable power into a single physically motivated metric. This interpretation is also consistent with figures of merit commonly used in resonant wireless power transfer systems, where coupling strength and Q-factor together determine the achievable power transfer.

\section{Data Anomalies, Repeatability, and Large-Signal Effects}\label{sec:SecIV}
\begin{figure} [bh!]
     \centering
     \subfloat[]
     %{\includegraphics[width=1\linewidth]{Figures/shrinkFlowChart.eps}}\\
   {\includegraphics[width=1\columnwidth]{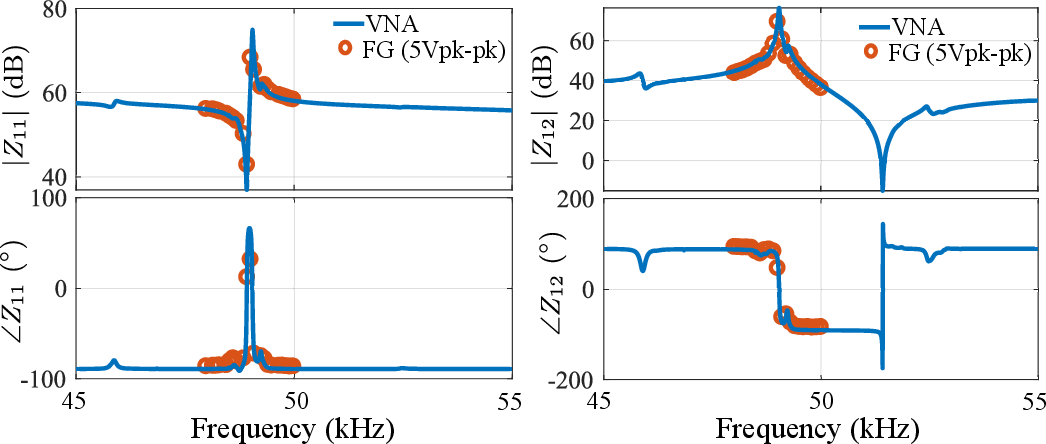}}  \\
   \subfloat[]
    {\includegraphics[width=1\linewidth]{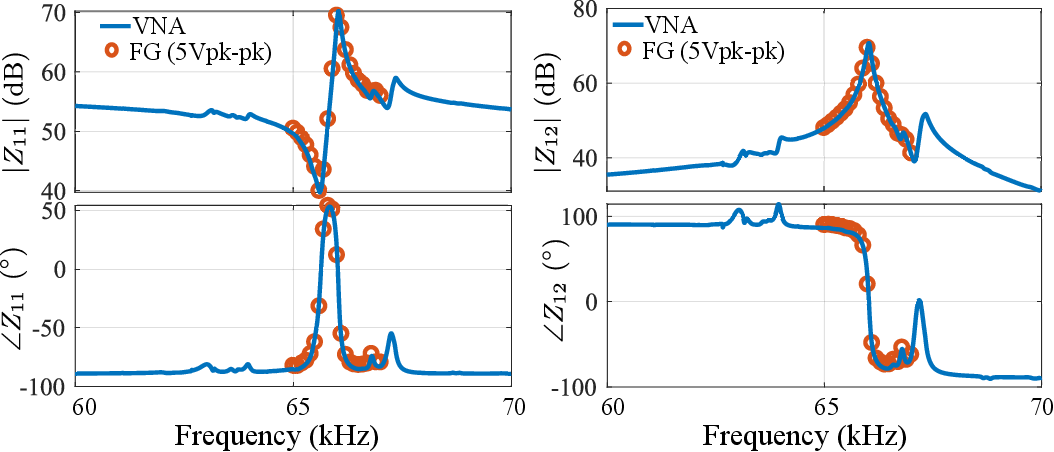}}\\
   %  \subfloat[]
    {\includegraphics[width=1\linewidth]{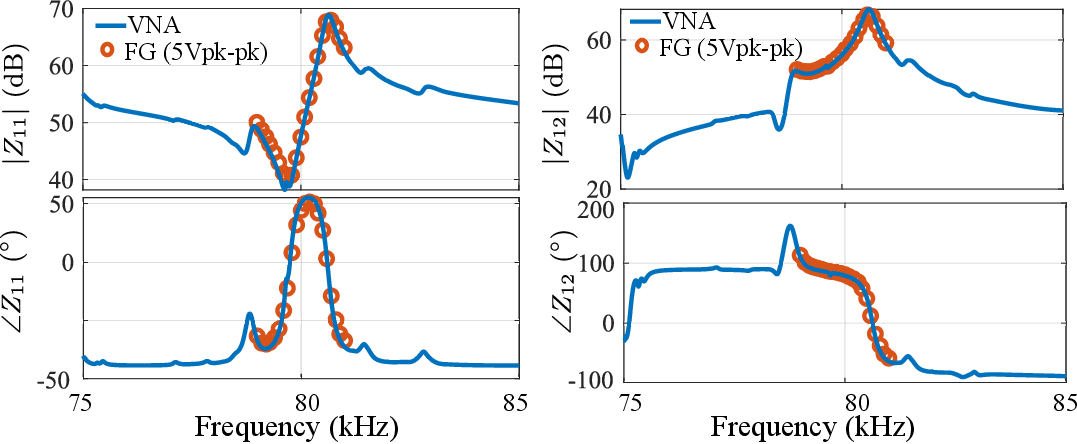}}
       \caption{Characteristics comparison for prototype~III for modes (a) 1, (b) 2, and (c) 3, from VNA data and Function Generator (FG) frequency sweep around the resonance frequency.} 
    \label{fig:repeatability}
\end{figure}
Based on the experimental characterization using a VNA, several practical factors have been identified that can introduce anomalies and reduce repeatability of the measured impedance characteristics, thereby affecting the accuracy of the predicted power transfer capability. These factors include proper VNA calibration, sufficient frequency resolution around high-Q resonances, reliable electrical connections to both Tx and Rx piezo electrodes, and consistent mechanical boundary conditions during measurement. In particular, placing the assembly on a rigid surface, wooden support, or mechanically soft material such as thermocol can alter the measured Q-factor. Therefore, repeated measurements and consistency checks are recommended before utilizing the extracted Z-parameters for power prediction. Fig.~\ref{fig:repeatability} demonstrates the characteristics measured around the resonance frequency of the steel-plate prototype using VNA and function generator excitation of 5 V (peak-to-peak) for the steel-plate prototype. It indicates good agreement and repeatability when these precautions are followed.

Furthermore, the power prediction presented in this work is based on small-signal VNA characterization, where the excitation voltage is typically a fraction of a volt. Consequently, the extracted Z-parameters represent the linear operating characteristics of the piezo–metal assembly. During practical operation, however, the transmitter piezo may be excited at several hundred volts, particularly in radial-mode operation, as indicated in Table~\ref{tab:OptPoints}. Under such conditions, nonlinear effects arising from the piezoelectric material, adhesive layer, and temperature-dependent property variations can modify the resonance frequencies and Q-factors. Similar voltage- and temperature-induced nonlinearities have been reported in the literature \cite{LargeSignalCh,TempNonlinear}.
Large-signal characterization can be performed by exciting the Tx and Rx piezos with higher-amplitude sinusoidal signals over the frequency range of interest and subsequently extracting the corresponding Z-parameters. Precautions must be taken near resonance frequencies, where high open-circuit voltages may develop across the Rx piezo, potentially damaging or degrading the piezoelectric elements. Detailed large-signal characterization procedures are available in the literature \cite{LargeSignalCh,Bode100} and are beyond the scope of the present work.
Nevertheless, the framework presented in Fig. \ref{fig:flowchart} for maximizing power transfer remains applicable irrespective of the excitation level. When large-signal Z-parameters are available, the same procedure can be employed to predict transferable power, determine optimal operating frequency, and identify the corresponding load conditions.

\section{Conclusion}
%This work presents an experimental investigation of power transfer capability in through-metal acoustic power transfer systems considering realistic piezo--metal assemblies. Significant deviations from standalone piezo behavior are observed due to bonding layers and assembly variations, leading to reduced Q-factor, frequency shifts, and spurious modes. A two-port network–based approach is employed to evaluate power transfer and optimal operating conditions under depolarization and thermal constraints. The results show that radial mode operation offers higher power transfer potential compared to thickness mode for the prototypes studied. A figure of merit (FOM) is proposed to enable rapid screening of configurations, demonstrating conssitency with the optimal power transfer obtained from detailed analysis. The presented approach provides practical design insights for achieving high-power operation in TM-APT systems.

This work experimentally evaluates power transfer capability in through-metal acoustic systems using realistic piezo–metal assemblies. Compared to standalone piezo behavior, bonding and assembly reduce Q-factor, shift resonance, and introduce spurious modes. A two-port network-based formulation is used to determine optimal operating conditions under depolarization and thermal constraints. Results show that, radial mode offers higher power transfer potential than thickness mode for the evaluated prototypes. A figure of merit (FOM) is proposed for rapid screening of prototypes, demonstrating close agreement with the optimal power transfer frequencies from detailed analysis. The proposed approach provides practical design and operational insights for high-power TM-APT systems while also highlighting the limitations of the methodology. The predicted optimal operating conditions provide useful guidelines for the design of the power electronic interface, compensation network, and control strategy.

\bibliographystyle{IEEEtran}

\bibliography{Bibliography}

\end{document}